\documentclass[acmnow]{acmtrans2m}

\usepackage[latin9]{inputenc} \usepackage[T1]{fontenc}

\usepackage[english]{babel}
\usepackage{amsmath}
\usepackage{amsfonts}
\usepackage{amssymb}
\usepackage{fancybox}
\usepackage{url}
\usepackage{graphicx}
\usepackage{xypic}
\usepackage{multicol}
\usepackage{rotating}
\usepackage{listings}

\def\runningfoot{\def\@runningfoot{.}}
\def\firstfoot{\def\@firstfoot{.}}

\begin{document}

\title{Certification of bounds on expressions involving rounded operators}
\markboth{Marc Daumas and Guillaume Melquiond}{Certification of bounds on expressions involving rounded operators}

\author{%
  Marc Daumas \\
  {\sc lirmm} ({\sc umr} 5506 {\sc cnrs}--{\sc um}2), {\rm visiting} {\sc éliaus} ({\sc éa 3679 upvd})
  \and
  Guillaume Melquiond \\
  {\sc lip (umr 5668 cnrs--éns} {\rm Lyon}--{\sc inria)}
}

\category{G.4}{Mathematical Software}{Certification and Testing}
\terms{Interval Arithmetic, Floating Point, Proof System}

\keywords{Forward error analysis, Dyadic fraction, Coq, PVS, HOL Light, Proof obligation}

\begin{abstract}
  Gappa uses interval arithmetic to certify bounds on mathematical
  expressions that involve rounded as well as exact operators.  Gappa
  generates a theorem with its proof for each bound treated.  The
  proof can be checked with a higher order logic automatic proof
  checker, either Coq or HOL Light, and we have developed a large
  companion library of verified facts for Coq dealing with the
  addition, multiplication, division, and square root, in fixed- and
  floating-point arithmetics.  Gappa uses multiple-precision dyadic
  fractions for the endpoints of intervals and performs forward error
  analysis on rounded operators when necessary.  When asked, Gappa
  reports the best bounds it is able to reach for a given expression
  in a given context.  This feature is used to quickly obtain
  coarse bounds. It can also be used to identify where the set of
  facts and automatic techniques implemented in Gappa becomes
  insufficient.  Gappa handles seamlessly additional properties
  expressed as interval properties or rewriting rules in
  order to establish more intricate bounds.  Recent work showed that
  Gappa is perfectly suited to the proof of correctness of small
  pieces of software. Proof obligations can be written by
  designers, produced by third-party tools or obtained by overloading
  arithmetic operators.
\end{abstract}

\maketitle
\newcommand{\R}{\mathbb{R}}
\newcommand{\Z}{\mathbb{Z}}
\newcommand{\IF}{\mathbb{IF}}
\renewcommand{\circ}{\texttt{rnd}}
\newcommand{\ulp}{\texttt{ulp}}
\newcommand{\circe}[1]{\texttt{err}_{\texttt{rnd},#1}}
\newcommand{\circa}{\texttt{err}_{\texttt{rnd}, \texttt{abs}}}
\newcommand{\circr}{\texttt{err}_{\texttt{rnd}, \texttt{rel}}}

\lstset{basicstyle=\tt\small,xleftmargin=1cm,numberstyle=\tiny}

\section{Introduction}

Gappa is a simple and efficient tool to certify bounds in computer
arithmetic~\cite{Rev06} and in the engineering of numerical
software~\cite{DinLauMel06,MelPio07,DauGio07} and hardware~\cite{MicTisVey06}.
Gappa bounds arithmetic expressions on real and rational numbers and
their evaluations in computers on fixed- and floating-point data
formats. Properties that are most often needed involve:
\begin{itemize}
\item ranges of rounded expressions to prevent exceptional behaviors
  (overflow, division by zero, and so on),
\item ranges of absolute and/or relative errors to characterize the
  accuracy of results. %
\end{itemize}

To the best of our knowledge, Gappa is a tool that was missing in
computer arithmetic and related research areas.  On one hand, Gappa is
not the first tool able to certify static ranges and error
bounds.  Two other projects are currently mixing interval arithmetic
and automatic proof checking \cite{GamMan04,DauMelMun05}. The first
one uses ACL2 \cite{KauManMoo2K} and the second one uses PVS
\cite{OwrRusSha92}.  Gappa is, however, the first tool able to certify
these bounds when the program relies on advanced numerical recipes
like error compensation, iterative refinement, {\em etc}.

On the other hand, countless efficient algorithms use symbolic
computation or interval arithmetic to produce bounds on expressions
but seldom provide gateways to automatic proof checkers. The
continuing work on interval arithmetic \cite{Neu90,JauKieDidWal01} has
created a huge set of useful techniques to deliver accurate answers in
a reasonable time.  Each technique is adapted to a specific class of
problems and most evaluations yield accurate bounds only if they are
handled by the appropriate techniques in the appropriate order.
Blending interval arithmetic and properties on dyadic fractions has
also been heavily used in computer arithmetic \cite{RumOgiOis05}.

Integration is certainly the challenge that prevented the development
of competitors to Gappa. Proofs generated by Gappa typically contain
4,800 of lines \cite{DinLauMel06} and related projects were not able
to avoid the development of small programs, for example to generate a
proof script about 9,935 intervals each requiring 3 theorems in PVS
\cite{DauMelMun05}.  The functionalities of Gappa presented here show
its potential in tackling generic problems that are unreachable with
other available tools. Our goal is to
\begin{itemize}
\item Provide {\em invisible formal methods} \cite{TiwShaRus03} in the
  sense that Gappa delivers formal certificates to users that are not
  expected to ever write any piece of proof in any formal proof
  system.
\item Provide a tool that is able to consider and combine many techniques using
  interval arithmetic, dyadic fractions, and rewriting rules.  Gappa
  performs an exhaustive search on its built-in set of facts and
  techniques. It is also able to follow hints given by users to take
  into account new techniques.
\item Simplify a valid proof once it has been produced in order to
  reduce the certification time, as in-depth proof checking is and
  will remain much slower than simple C++ evaluation.
\item Provide a tool appropriate to meet the highest Common Criteria
  Evaluated Assurance Level (EAL~7) \cite{Sch03,Roc05} for numerical
  applications using floating- and fixed-point arithmetics.
\end{itemize}

Gappa is composed of two parts. First, a program written in C++, based
on Boost interval arithmetic library~\cite{BroMelPio03} and
MPFR~\cite{FouHanLefPelZim05}, verifies numeric properties given by the
user. Along these verifications, it generates formal certificates of
their validity. Second, a companion library provides theorems with
computable hypotheses. This set of theorems allows a proof assistant to
interpret the formal certificates and hence to automatically check the
validity of the numeric properties. The proof assistant we use is
Coq~\cite{HueKahPau04}, but ongoing work shows that Gappa can generate
formal certificates for other proof assistants such as HOL
Light~\cite{Har2Kb}.

We first describe the input language of Gappa and we detail its built-in
rewriting rules. We then present the set of theorems and interval
operators Gappa relies on to prove numeric properties and we describe
how it interacts with proof checkers, extending \cite{DauMel04}. We
finish this report with perspectives, experiments, and concluding
remarks.

\section{Description of the input language of Gappa}

Consider for example that $y$ is the result of a portion of code
without loops and branches. The definition of $y$ is an expression
involving rounded operators and rounded constants.  We may define $Y$
(uppercase) as the exact answer without any rounding error. The
expression $Y$ is identical to $y$ except that rounded operators are
replaced by exact operators and rounded constants are replaced by
exact constants. If some numeric terms were considered negligible and
were optimized out of the implementation $y$, these terms are
introduced in $Y$.  So the expression $y$ gives the effectively
computed value while the expression $Y$ gives the ideal value $y$
tries to approximate.

In order to certify the correctness of this code, we will possibly need
\begin{itemize}
\item an interval containing all the possible values of $y$ to guarantee
  that $y$ does not overflow and produces no invalid value,
\item an interval containing all the possible values of $y - Y$ or $(y
  - Y)/Y$ to guarantee that $y$ is accurate and close to $Y$.
\end{itemize}

The grammar of the input language to Gappa is presented in
Figure~\ref{fig/gram}. It has been designed to efficiently express
such needs. An input file is composed of three parts: a set of aliases
({\tt PROG}, detailed in Section~\ref{sub/prog}), the proposition to
be proved ({\tt PROP}, detailed in Section~\ref{sub/prop}) and a set
of hints ({\tt HINTS}, detailed in Sections~\ref{sub/hints} and
\ref{sub/subp}).  When successful, Gappa produces a Coq or a HOL Light
file with the proof of {\tt PROP}. Its validity can be checked by Coq
using the companion library and by HOL Light using a set of axioms
until a companion library becomes available.

\begin{sidewaysfigure}
\begin{center}
\begin{multicols}{2}
\small
\begin{verbatim}
 0 $accept: BLOB $end
 1 BLOB: PROG '{' PROP '}' HINTS
 2 PROP: REAL LE SNUMBER
 3     | REAL IN '[' SNUMBER ',' SNUMBER ']'
 4     | REAL IN '?'
 5     | REAL GE SNUMBER
 6     | PROP AND PROP
 7     | PROP OR PROP
 8     | PROP IMPL PROP
 9     | NOT PROP
10     | '(' PROP ')'
11 SNUMBER: NUMBER
12        | '+' NUMBER
13        | '-' NUMBER
14 FUNCTION_PARAM: SNUMBER
15               | IDENT
16 FUNCTION_PARAMS_AUX: FUNCTION_PARAM
17                    | FUNCTION_PARAMS_AUX ',' FUNCTION_PARAM
18 FUNCTION_PARAMS: /* empty */
19                | '<' FUNCTION_PARAMS_AUX '>'
20 FUNCTION: IDENT FUNCTION_PARAMS
21 EQUAL: '='
22      | FUNCTION '='
23 PROG: /* empty */
24     | PROG IDENT EQUAL REAL ';'
25     | PROG '@' IDENT '=' FUNCTION ';'


          26 REAL: SNUMBER
          27     | IDENT
          28     | FUNCTION '(' REALS ')'
          29     | REAL '+' REAL
          30     | REAL '-' REAL
          31     | REAL '*' REAL
          32     | REAL '/' REAL
          33     | '|' REAL '|'
          34     | SQRT '(' REAL ')'
          35     | FMA '(' REAL ',' REAL ',' REAL ')'
          36     | '(' REAL ')'
          37     | '+' REAL
          38     | '-' REAL
          39 REALS: REAL
          40      | REALS ',' REAL
          41 DPOINTS: SNUMBER
          42        | DPOINTS ',' SNUMBER
          43 DVAR: REAL
          44     | REAL IN NUMBER
          45     | REAL IN '(' DPOINTS ')'
          46 DVARS: DVAR
          47      | DVARS ',' DVAR
          48 HINTS: /* empty */
          49      | HINTS REAL IMPL REAL ';'
          50      | HINTS REALS '$' DVARS ';'
          51      | HINTS '$' DVARS ';'
          52      | HINTS REAL '~' REAL ';'
\end{verbatim}
    \end{multicols}
    \caption{Grammar of the input language to Gappa generated by {\tt byson}}
    \label{fig/gram}
  \end{center}
\end{sidewaysfigure}

\subsection{Formalizing the proposition ({\tt PROP}) that Gappa proves}
\label{sub/prop}

The proposition ({\tt PROP}) that Gappa is expected to prove is
written between brackets~({\tt \{ \}}) as presented below and it may
contain any conjunction ({\tt AND} token: \verb+/\+),
disjunction~({\tt OR} token: \verb+\/+), implication ({\tt IMPL}
token: {\tt ->}) or negation ({\tt NOT} token: {\tt not}) of
enclosures of expressions.  Enclosures are either inequalities ({\tt
  LE} or {\tt GE} tokens: {\tt <=} or {\tt >=}) or bounded ranges
({\tt IN} token: {\tt in}) on expressions ({\tt REAL}).  Ranges may be
left unspecified by using question marks (\verb+?+) instead of
intervals.  Endpoints of intervals and bounds of inequalities are
numerical constants ({\tt SNUMBER}). 
\begin{lstlisting}
{ x - 2 in [-2,0] /\ (x + 1 in [0,2] -> y in [3,4])
  -> not x <= 1 \/ x + y in ? }
\end{lstlisting}
Expressions ({\tt REAL}) may contain constants ({\tt SNUMBER}),
identifiers ({\tt IDENT}), user-defined as well as built-in rounding
operators ({\tt FUNCTION}), and arithmetic operators (addition,
subtraction, multiplication, division, absolute value, square root,
negation, and fused multiply and add).

The goal of Gappa is to prove the whole logical proposition, assuming
that undefined identifiers ($x$ and $y$ in the example above) are
universally quantified over the set of real numbers. If question marks
are used in some expression enclosures, Gappa suggests intervals for these
enclosures such that the proposition can be proved. In the example
above, Gappa suggests $x + y \in [3,5]$, which happens to be the
tightest interval such that the proposition holds true. Question marks
are not allowed if they induce unspecified hypotheses once
transformations of Section~\ref{sub/work-on-prop} have been applied.

As Gappa stores interval endpoints as dyadic fractions, it produces an
error message when a goal contains an interval so tight that it has to
be replaced with an empty interval. For example, Gappa is unable to
prove the goal {\tt 13/10 in [1.3,1.3]}, as the empty set is the biggest
representable subset of the set $\{1.3\}$.

The fact that bounds are numerical constants is not a strong
limitation to the use of Gappa. For example, linear dependencies on
intervals can be introduced by manipulating expressions: the
enclosure $y - Y \in [-i\times 10^{-6}, i\times 10^{-6}]$ is not
allowed, but the enclosure $(y - Y)/i \in [-10^{-6}, 10^{-6}]$ is.

\subsection{Definitions of aliases to describe the behavior of programs ({\tt PROG})}
\label{sub/prog}

Typing large expressions in the proposition ({\tt PROP} seen
Section~\ref{sub/prop}) would not be practical for proof obligations
generated from actual pieces of software. Aliases ({\tt IDENT}) of
expressions ({\tt REAL}) are defined by constructions of the form
{\tt IDENT = REAL}. {\tt IDENT} becomes available for later
definitions, the proposition, and the hints. This construction is neither
an equality nor an affectation but rather an alias.  Gappa uses {\tt
  IDENT} for its outputs and in the formal proof instead of machine
generated names. An identifier cannot be aliased more than once, even
if the right hand sides of both aliases are equivalent. Neither can it
be aliased after having been used as an unbound variable. For example
{\tt b = a * 2; a = 1;} is not allowed.

Rounding operators are used in the arithmetic expressions describing
the behavior of numerical codes. They are real functions yielding
rounded values according to the target data format (\texttt{precision}
and \texttt{minimum\_exponent}, or \texttt{lsb\_weight}) and a
predefined rounding mode amongst the ones presented
Table~\ref{tab/dir}. For modes that are not defined by IEEE 754
standard \cite{Ste.87} and its forthcoming revision, see
\cite{EveSei99,BolMel05} and references herein.  Floating- and
fixed-point rounding operators can be expressed with the following
operators where rounding parameters ({\tt FUNCTION\_PARAMS}) are
listed between angle brackets:
\begin{lstlisting}[xleftmargin=0cm]
float< precision, minimum_exponent, rounding_direction >(...)
fixed< lsb_weight, rounding_direction >(...)
\end{lstlisting}

The syntax above can be abbreviated for the floating-point formats of
Table~\ref{tab/for} and for (fixed-point) integer arithmetic:
\begin{lstlisting}[xleftmargin=0cm]
float< name, rounding_direction >(...)
int< rounding_direction >(...)
\end{lstlisting}

Aliases are permitted for rounding operators. Their definitions are
prefixed by the '@' sign. Line 1 below defines the {\tt rnd} function
as rounding to nearest using IEEE 754 standard for 32 bit
floating-point data. The example shows various
ways of expressing rounded operations using the alternate constructions
of {\tt EQUAL}.  When all the arithmetic
operations on the right hand side of an alias are followed by the same
rounding operator (as visible Line 2), this operator can be put once
and for all on the left of the equal symbol (as presented Line 3).  On
this example, Gappa even complains that $y$ and $z$ are two
different names for the same expression.
\begin{lstlisting}[numbers=left]
@rnd = float< ieee_32, ne>;
y = rnd(x * rnd(1 - x));
z rnd= x * (1 - x);
\end{lstlisting}

\begin{table}[htbp]
  \caption{Rounding modes available in Gappa}
  \label{tab/dir}
  \begin{center}
    \begin{tabular}{|c|l|c|} \hline
      Alias     & Meaning                                        \\ \hline
      \tt zr    & toward zero                                    \\
      \tt aw    & away from zero                                 \\
      \tt dn    & toward minus infinity (down)                   \\
      \tt up    & toward plus infinity                           \\
      \tt od    & to odd mantissas                               \\
      \tt ne    & to nearest, tie breaking to even mantissas     \\
      \tt no    & to nearest, tie breaking to odd mantissas      \\
      \tt nz    & to nearest, tie breaking toward zero           \\
      \tt na    & to nearest, tie breaking away from zero        \\
      \tt nd    & to nearest, tie breaking toward minus infinity \\
      \tt nu    & to nearest, tie breaking toward plus infinity  \\ \hline
    \end{tabular}
  \end{center}
\end{table}

\begin{table}[htbp]
  \caption{Predefined floating-point formats available in Gappa}
  \label{tab/for}
  \begin{center}
    \begin{tabular}{|c|l|c} \hline
      Alias         & Meaning                      \\ \hline
      \tt ieee\_32  & IEEE-754 single precision    \\
      \tt ieee\_64  & IEEE-754 double precision    \\
      \tt ieee\_128 & IEEE-754 quadruple precision \\
      \tt x86\_80   & 80 bit extended precision    \\ \hline
    \end{tabular}
  \end{center}
\end{table}

Most truncated hardware operators \cite{Tex97} and some compound
operators cannot be described as if they were first computed to infinite
precision and then rounded to target precision. For such operators we revert to
under-specified functions that produce results with a known bound on
the relative error.
\begin{lstlisting}[xleftmargin=0cm]
{add|sub|mul}_rel< precision [, minimum_exponent] >(..., ...)
\end{lstlisting}
If a minimum exponent is provided, Gappa does not instantiate any
assumption that involves a result with an exponent below the minimum
exponent.  Otherwise, the error bound always holds and the absolute
error is 0 when the result is 0.

\subsection{Rewriting expressions to suppress some dependency effects (first use of {\tt HINT})}
\label{sub/hints}

Let $Y$ be an expression and $y$ an approximation of $Y$ due to
round-off errors, for example.  The absolute error is $y - Y$ and the
relative error is $(y - Y) / Y$.  As soon as Gappa has computed some ranges
for $y$ and $Y$, it naively computes an enclosing interval of $y - Y$
and $(y - Y) / Y$ using theorems on subtraction and division of
intervals.

Unfortunately, expressions $y$ and $Y$ are strongly correlated and
error ranges computed that way are useless. To suppress some
dependency effects and reproduce many of the techniques used in
numerical analysis and in computer
arithmetic~\cite{Kah65,Hig02,BolDau04b,DinDefLau04}, Gappa manipulates
error expressions through a set of built-in pattern-matching as well
as user-defined rewriting rules.

We assume that $y = \circ(a + b)$ and $Y = A + B$. Gappa rewrites
the absolute error $\circ(a + b) - (A + B)$ as $(\circ(a + b) - (a +
b)) + ((a + b) - (A + B))$. It finds an enclosure of the first term
using a theorem on the $\circ$ rounding operator. For the second term,
Gappa performs a second rewrite: $(a + b) - (A + B)$ is equal to $(a -
A) + (b - B)$. This rewriting rule gives sensible results, as long as
$a$ and $b$ are close to $A$ and $B$ respectively.

\begin{table}
  \caption{Built-in rewriting rules available in Gappa}
  \label{tab/rules}
  $$
  \begin{array}{| c | c | c | c |} \hline
    \text{Rule}         & \text{Before}                    & \text{After}                                    & \text{Condition}           \\ \hline
    \texttt{opp\_mibs}    & -a - -b                          & -(a - b)                                        & a   \neq b                               \\
    \texttt{opp\_mibs}    & (-a - -b) / -b                   &  (a - b) / b                                    & b   \neq 0 \land  a \neq b               \\
    \texttt{add\_xals}    & a + b                            & (a - A) + (A + b)                               &                                          \\
    \texttt{add\_xars}    & c + a                            & (c + A) + (a - A))                              &                                          \\
    \texttt{add\_mibs}    & (a + b) - (c + d)                & (a - c) + (b - d)                               & a   \neq c \land  b \neq d               \\
    \texttt{add\_fils}    & (a + b) - (a + c)                & b - c                                           & b   \neq c                               \\
    \texttt{add\_firs}    & (a + b) - (c + b)                & a - c                                           & a   \neq c                               \\
    \texttt{sub\_xals}    & a - b                            & (a - A) + (A - b)                               & a   \neq b \land A \neq b                \\
    \texttt{sub\_xars}    & b - a                            & (b - A) + -(a - A)                              & b   \neq a                               \\
    \texttt{sub\_mibs}    & (a - b) - (c - d)                & (a - c) + -(b - d)                              & a   \neq c \land b \neq d                \\
    \texttt{sub\_fils}    & (a - b) - (a - c)                & -(b - c)                                        & b   \neq c                               \\
    \texttt{sub\_firs}    & (a - b) - (c - b)                & a - c                                           & a   \neq c                               \\
    \texttt{mul\_xals}    & a b                              & (a - A) b + A b                                 &                                          \\
    \texttt{mul\_xars}    & b a                              & b (a - A) + b A                                 &                                          \\
    \texttt{mul\_fils}    & a b - a c                        & a (b - c)                                       & b   \neq c                               \\
    \texttt{mul\_firs}    & a c - b c                        & (a - b) c                                       & a   \neq b                               \\
    \texttt{mul\_mars}    & a b - c d                        & a (b - d) + (a - c) d                           & a   \neq c \land b \neq d                \\
    \texttt{mul\_mals}    & a b - c d                        & (a - c) b + c (b - d)                           & a   \neq c \land b \neq d                \\
    \texttt{mul\_mabs}    & a b - c d                        & a (b - d) + (a - c) b + -((a - c) (b - d))      & a   \neq c \land b \neq d                \\
    \texttt{mul\_mibs}    & a b - c d                        & c (b - d) + (a - c) d + (a - c) (b - d)         & a   \neq c \land b \neq d                \\
    \texttt{mul\_filq}    & (a b - a c) / (a c)              & (b - c) / c                                     & ac  \neq 0 \land b \neq c                \\
    \texttt{mul\_firq}    & (a b - c b) / (c b)              & (a - c) / c                                     & bc  \neq 0 \land a \neq c                \\
    \texttt{div\_mibq}    & (a / b - c / d) / (c / d)        & ((a - c) / c - (b - d) / d) / (1 + (b - d) / d) & bcd \neq 0 \land b \neq d                \\
    \texttt{div\_firq}    & (a / b - c / b) / (c / b)        & (a - c) / c                                     & bc  \neq 0 \land a \neq c                \\
    \texttt{sqrt\_mibs}   & \sqrt{a} - \sqrt{b}              & (a - b) / (\sqrt{a} + \sqrt{b})                 & a   \ge  0 \land b \ge  0 \land a \neq b \\            
    \texttt{sqrt\_mibq}   & (\sqrt{a} - \sqrt{b}) / \sqrt{b} & \sqrt{1 + (a - b) / b} - 1                      & a   \ge  0 \land b >    0 \land a \neq b \\
    \texttt{sub\_xals}    & b - A                            & (b - a) + (a - A)                               & A   \neq b \land a \neq b                \\
    \texttt{err\_fabq}    & 1 + (a - b) / b                  & a / b                                           & b   \neq 0 \land a \neq b                \\
    \texttt{val\_xabs}    & a                                & A + (a - A)                                     &                                          \\
    \texttt{val\_xebs}    & A                                & a + -(a - A)                                    &                                          \\
    \texttt{val\_xabq}    & a                                & A (1 + (a - A) / A)                             & a   \neq 0                               \\
    \texttt{val\_xebq}    & A                                & a / (1 + (a - A) / A)                           & ab  \neq 0                               \\
    \texttt{square\_sqrt} & \sqrt{a} \times \sqrt{a}         & a                                               & a \ge 0                                  \\
    \texttt{addf\_1}      & a / (a + b)                      & 1 / (1 + b / a)                                 & a(a + b) \neq 0 \land  a \neq 1          \\
    \texttt{addf\_2}      & a / (a + b)                      & 1 - 1 / (1 + a / b)                             & b(a + b) \neq 0 \land a \neq 1           \\
    \texttt{addf\_3}      & a / (a - b)                      & 1 / (1 - b / a)                                 & a(a - b) \neq 0 \land a \neq 1           \\
    \texttt{addf\_4}      & a / (a - b)                      & 1 + 1 / (a / b - 1)                             & b(a - b) \neq 0 \land a \neq 1           \\ \hline
  \end{array}
  $$
\end{table}

Table~\ref{tab/rules} contains some of the rules Gappa 
tries to apply automatically. There are two kinds of rewriting rules. Rules of
the first kind, for example {\tt add\_firs}, are meant to produce
simpler expressions. Rules of the second kind, for example {\tt
  sub\_xals}, are used to reproduce common practices of computer
arithmetic by introducing intermediate terms in expressions.  In order
for an expression to match an uppercase letter in such a rule, the expression
that matches the same letter in lowercase has to be tagged as an
approximation of the former.

The first rule, {\tt sub\_xals}, has been applied earlier by Gappa,
because Gappa automatically tags $\circ(x)$ as an approximation of
$x$, for any expression $x$, since $\circ$ is a rounding operator.
Gappa also creates such pairs for expressions that define absolute and
relative errors in some hypotheses of a sub-formula of the proposition
{\tt PROP}. For example, on the following input, Gappa proposes
accurate bounds, as it considers $x$ to be an approximation of $y$,
and $\lfloor x \rfloor$ of $x$.
\begin{lstlisting}
@floor = int<dn>;
{ x - y in [-0.1,0.1] -> floor(x) - y in ? }
\end{lstlisting}

Thanks to its built-in rewriting rules and its heuristics to detect
approximations, Gappa is able to automatically verify most properties on
numerical applications that use common practices. Gappa, however, is not
a complete decision procedure\footnote{While seemingly simple, the
formalism of Gappa is rich enough so that any first-order formula for
Peano arithmetic can be expressed. As a consequence, it is impossible to
design an algorithm that is able to automatically decide whether any
proposition is provable or not.} and it may fail to prove some
propositions. When that happens, users can give some hints to the tool.

For example, in the above input, a user could add the hint \verb|x ~ y|
after the proposition, in order to indicate that $x$ is an approximation
of $y$. As Gappa was able to guess this property automatically, Gappa
will warn that the hint is useless: the user can remove it.

Another kind of hint allows the users to directly add rewriting
rules. The hint \texttt{primary -> secondary} states that Gappa
can use an enclosure of {\tt secondary} expression whenever it needs
an enclosure of {\tt primary} expression. For example, the following
hint describes Newton's relation between the reciprocal $\frac{1}{y}$
and its approximation $x \cdot (2 - x \cdot y)$.
\begin{lstlisting}
x * (2 - x * y) - 1/y -> (x - 1/y) * (x - 1/y) * -y
\end{lstlisting}

Such rules usually explicit
some techniques applied by designers that are not necessarily visible
in the source code.  We cannot expect an automatic tool to re-discover
innovative techniques. Yet, we will incorporate in Gappa any technique
that we find to be commonly used. Any additional rewriting rule produces
an hypothesis in the generated Coq file that must be proved
independently, for example with the \texttt{ring} tactic of Coq.

In order for the \texttt{primary -> secondary} rule to be valid, any
value of {\tt primary} must be contained in the computed enclosure of
{\tt secondary}. This property generally holds true if both
expressions are equal. As a consequence, Gappa tries to check if they
are equal and warns if they are not, in order to detect mistypings
early. Note that Gappa does not check if divisors are always different
from zero before applying user-defined rewriting rules. Yet, Gappa
detects divisors that are trivially equal to zero in expressions that
appear in rewriting rules.  For example, \texttt{y -> y * (x - x) / (x
  - x)} is most certainly an error.

Due to built-in and user-defined rewriting rules, Gappa may hold more
than one expression for a quantity, and hence several bounds as
evaluations of equivalent expressions in interval arithmetic may yield
different results. The intersection of the intervals yielded by the
different expressions may be tighter than its previously known bounds.
Tightening bounds on one quantity may then lead to tighter bounds on
quantities based on it.  Gappa explores the directed acyclic graph of
quantities breadth-first until its goal is achieved or all bounds of
the graph stopped evolving.

\subsection{Sub-paving the range of some quantities by bisection (second use of {\tt HINT})}
\label{sub/subp}

The last kind of hint that can be used when Gappa is unable to prove
automatically a formula is to pave the range of some quantities and to
prove independent results on each tile.  Rewriting expressions is
usually very efficient but it fails if different proof structures are
needed on various parts of the range, as in the following example. The
generic proof structure only works for $x \in [0,\frac{1}{2}]$. A
specific proof structure is needed in order to extend the result to $x
\in [\frac{1}{2},3]$. This proof relies on the fact that $\circ(y) -
y$ is always zero there. But Gappa will not notice this property
unless the last line is provided.
\begin{lstlisting}
@rnd = float< ieee_32, ne >;
x = rnd(x_);
y = x - 1;
z = x * (rnd(y) - y);
{ x in [0,3] -> |z| <= 1b-26 }
|z| $ x;
\end{lstlisting} %
There are three constructions for bisection each involving a {\tt \$}
sign in the hints section:
\begin{itemize}
\item Evenly split the range into as many sub-intervals as asked.
  E.g. \texttt{\$ x in 6} splits the range of $x$ in six sub-intervals.
  If the number of intervals is omitted (e.g. \texttt{\$~x}) and no expression
  is present on the left of {\tt \$}, the default is 4.
\item Split an interval on user-provided points. E.g. \texttt{\$ x in
    (0.5,2)} splits the range $[0,3]$ of $x$ above in three
  sub-intervals, the middle one being $[0.5,2]$.
\item The third kind of bisection tries to find by dichotomy a good
  sub-paving such that one goal of the proposition holds. The
  range of this goal has to be specified in the proposition, and the
  concerned expression has to be indicated on the left of the {\tt \$}
  symbol.
\end{itemize}

More than one bisection hint can be used and hints of the third kind
can try to satisfy more than one goal at once. The two hints below
will be used sequentially one after the other. The first one splits
the range of $u$ until all the enclosures on $a$, $b$, and $c$ are
verified.
\begin{lstlisting}
a, b, c $ u;
d, e $ v;
\end{lstlisting}

Users may build higher dimension sub-paving by using more than one term
on the right of the {\tt \$} symbol, reaching quickly combinatorial
explosions though.  The following statement asks Gappa to find a set of
sub-ranges of $u$ and $w$ such that the goals on $a$ and $b$ are
satisfied when the range of $v$ is split into three sub-intervals.
\begin{lstlisting}
a, b $ u, v in 3, w
\end{lstlisting} %

\section{Handling automatic proof checkers}

\subsection{Work on the logical proposition ({\tt PROP})}
\label{sub/work-on-prop}

The proposition is first modified and loosely broken according to the
rules of sequent calculus as presented below for the proposition seen
in Section~\ref{sub/prop}.
\begin{lstlisting}
{ x - 2 in [-2,0] /\ (x + 1 in [0,2] -> y in [3,4])
  -> not x <= 1 \/ x + y in ? }
\end{lstlisting}
Each of the following new formulas is then verified by Gappa. If both
formulas hold true, the original proposition does too.
\begin{eqnarray*}
x \le 1 \land x - 2 \in [-2,0] &\Longrightarrow& x + 1 \in [0,2] \lor x + y \in \texttt{?}\\
x \le 1 \land x - 2 \in [-2,0] \land y \in [3,4] &\Longrightarrow& x + y \in \texttt{?}
\end{eqnarray*}

Gappa performs this decomposition in order to obtain implication
formulas with conjunctions of enclosures on their left hand sides and
trees of conjunctions and disjunctions of enclosures on their right hand
sides. In particular, all the negation symbols and the inner
implications have been removed. For example, a set of implications of
the form $e_1 \in I_1 \land \cdots \land e_m \in I_m \Rightarrow f_1 \in
J_1 \lor \cdots \lor f_m \in J_m$ is suitable for a use by Gappa.
Unspecified ranges (interrogation marks) are allowed as long as they
appear only on the right hand sides of these decomposed formulas.

Inequalities may appear on both sides of the implications. Any
inequality on the left hand side will be used only if Gappa can compute
an enclosure of the expression by some other means. Any inequality on
the right hand side is copied to the hypotheses as permitted by
classical logic, provided that it is reverted first. For example,
proposition $x \in [2,3] \Rightarrow (y \in [4,5] \land z \ge 6)$ is
equivalent to proposition $(x \in [2,3] \land z \le 6) \Rightarrow (y
\in [4,5] \land z \ge 6)$, but the second one provides a bigger set of
usable enclosures on its left hand side.

When the right hand side of the formula is a disjunction, Gappa searches
for a sub-term that holds under the hypotheses of the proposition. It
fails to prove valid disjunctions if it cannot find one sub-term that
always holds under the hypotheses.

\subsection{Structure of the generated proof}

\begin{table}
  \caption{Theorems on interval arithmetic available from the Coq companion library to Gappa}
  \label{tab/lib}
  $$
  \begin{array}{| l | l | l |} \hline
    \text{Target}                                      & \text{Hypotheses}                                                              & \text{Constraint}                 \\ \hline
    \mathtt{BND}(\circ(a) - a, I)                      &                                                                                & I \supset \circe{0} \\
    \mathtt{BND}(\circ(a) - a, I)                      & \mathtt{BND}(a, J)                                                             & I \supset \circe{1}(J) \\
    \mathtt{BND}(\circ(a) - a, I)                      & \mathtt{BND}(\circ(a), J)                                                      & I \supset \circe{2}(J) \\
    \mathtt{BND}(\circ(a) - a, I)                      & \mathtt{ABS}(a, J)                                                             & I \supset \circe{3}(J) \\
    \mathtt{BND}(\circ(a) - a, I)                      & \mathtt{ABS}(\circ(a), J)                                                      & I \supset \circe{4}(J) \\
    \mathtt{BND}((\circ(a) - a)/a, I)                  & \mathtt{BND}(a, J)                                                             & I \supset \circe{5}(J) \\
    \mathtt{BND}((\circ(a) - a)/a, I)                  & \mathtt{BND}(\circ(a), J)                                                      & I \supset \circe{6}(J) \\
    \mathtt{BND}((\circ(a) - a)/a, I)                  & \mathtt{ABS}(a, J)                                                             & I \supset \circe{7}(J) \\
    \mathtt{BND}((\circ(a) - a)/a, I)                  & \mathtt{ABS}(\circ(a), J)                                                      & I \supset \circe{8}(J) \\
    \mathtt{BND}(\circ(a), I)                          & \mathtt{BND}(a, J)                                                             & I \supset \circ(J) \\
    \mathtt{BND}(\circ(a), I)                          & \mathtt{BND}(\circ(a), J)                                                      & I \supset J \cap \mathbb{F}_{\circ} \\
    \mathtt{BND}(-a, I)                                & \mathtt{BND}(a, J)                                                             & I \supset -J \\
    \mathtt{BND}(|a|, I)                               & \mathtt{BND}(a, J)                                                             & I \supset |J| \\
    \mathtt{BND}(\sqrt{a}, I)                          & \mathtt{BND}(a, J)                                                             & J \ge 0 \land I \supset \sqrt{J} \\
    \mathtt{BND}(a - a, I)                             &                                                                                & 0 \in I  \\
    \mathtt{BND}(a/a, I)                               & \mathtt{ABS}(a, J)                                                             & 1 \in I \land J > 0 \\
    \mathtt{BND}(a \times a, I)                        & \mathtt{BND}(a, J)                                                             & I \supset |J| \times |J|\\
    \mathtt{BND}(a + b, I)                             & \mathtt{BND}(a, J), \mathtt{BND}(b, K)                                         & I \supset J + K \\
    \mathtt{BND}(a - b, I)                             & \mathtt{BND}(a, J), \mathtt{BND}(b, K)                                         & I \supset J - K \\
    \mathtt{BND}(a \times b, I)                        & \mathtt{BND}(a, J), \mathtt{BND}(b, K)                                         & I \supset J K \\
    \mathtt{BND}(a/b, I)                               & \mathtt{BND}(a, J), \mathtt{BND}(b, K)                                         & 0 \not\in K \land I \supset J / K \\
    \mathtt{ABS}(-a, I)                                & \mathtt{ABS}(a, J)                                                             & I \supset J \\
    \mathtt{ABS}(|a|, I)                               & \mathtt{ABS}(a, J)                                                             & I \supset J \\
    \mathtt{ABS}(\sqrt{a}, I)                          & \mathtt{ABS}(a, J)                                                             & I \supset \sqrt{J} \\
    \mathtt{ABS}(a \pm b, I)                           & \mathtt{ABS}(a, J), \mathtt{ABS}(b, K)                                         & I \supset |J - K| \cup (J + K) \\
    \mathtt{ABS}(a \times b, I)                        & \mathtt{ABS}(a, J), \mathtt{ABS}(b, K)                                         & I \supset J \times K \\
    \mathtt{ABS}(a/b, I)                               & \mathtt{ABS}(a, J), \mathtt{ABS}(b, K)                                         & K > 0 \land I \supset J / K \\
    \mathtt{BND}(a, I)                                 & \mathtt{ABS}(a, J)                                                             & I \supset J \cup -J \\
    \mathtt{BND}(a, I)                                 & \mathtt{BND}(a, J), \mathtt{ABS}(a, K)                                         & I \supset (J \cap K) \cup (J \cap -K) \\
    \mathtt{BND}(|a|, I)                               & \mathtt{ABS}(a, J)                                                             & I \supset J \\
    \mathtt{ABS}(a, I)                                 & \mathtt{BND}(|a|, J)                                                           & I \supset J \\
    \mathtt{BND}(a + b + a \times b, I)                & \mathtt{BND}(a, J), \mathtt{BND}(b, K)                                         & J \ge -1 \land K \ge -1 \land I \supset \mathcal{U}(J,K) \\
    \mathtt{BND}(\xi, I)                               &                                                                                & I \supset \{\xi\} \\
    \mathtt{FIX}(a \pm b, e)                           & \mathtt{FIX}(a, f), \mathtt{FIX}(b, g)                                         & e \le \min(f, g) \\
    \mathtt{FIX}(a \times b, e)                        & \mathtt{FIX}(a, f), \mathtt{FIX}(b, g)                                         & e \le f + g \\
    \mathtt{FLT}(a \times b, p)                        & \mathtt{FLT}(a, q), \mathtt{FLT}(b, r)                                         & p \ge q + r \\
    \mathtt{FIX}(a, e)                                 & \mathtt{FLT}(a, q), \mathtt{ABS}(a, J)                                         & J > 0 \land e \le 1 + \log_2(\underline{J}) - q \\
    \mathtt{FLT}(a, p)                                 & \mathtt{FIX}(a, e), \mathtt{ABS}(a, J)                                         & p \ge 1 + \log_2(\overline{J}) - e\\
    \mathtt{FIX}(a, e)                                 & \mathtt{BND}(a, [x,x])                                                         & \exists m \in \mathbb{Z}, x = m \cdot 2^e \\
    \mathtt{FLT}(a, p)                                 & \mathtt{BND}(a, [x,x])                                                         & \exists m,e \in \mathbb{Z}, x = m \cdot 2^e \land |m| < 2^p \\
    \mathtt{FIX}(\circ(a), e)                          &                                                                                & e \le e_\circ \\
    \mathtt{FLT}(\circ(a), p)                          &                                                                                & p \ge p_\circ \\
    \mathtt{BND}(\circ(a) - a, I)                      & \mathtt{FIX}(a, e), \mathtt{FLT}(a, p)                                         & 0 \in I \land e \ge e_\circ \land p \le p_\circ \\ \hline
  \end{array}
  $$
\end{table}

\begin{table}
  \caption{Interval operators used in Table~\ref{tab/lib}}
  \label{tab/int}
  $$
  \begin{array}{| l | l | l |} \hline
    \text{Operation}           & \text{Constraint} & \text{Definition}                                                                                                  \\ \hline
                               &                   &                                                                                                                    \\[-8pt]
    -I                         &                   & [- \overline{I}               ,  -\underline{J}  ]                                                                 \\
    I^{-1}                     & 0 \not\in I       & [1/\overline{I}, 1/\underline{I}]                                                                                  \\
    I + J                      &                   & [\underline{I} + \underline{J},  \overline{I} + \overline{J}   ]                                                   \\
    I - J                      &                   & I + (-J)                                                                                                           \\
    I \times J                 &                   & [\min(\underline{I}\underline{J}, \underline{I}\overline{J}, \overline{I}\underline{J}, \overline{I}\overline{J}),
                                                      \max(\underline{I}\underline{J}, \underline{I}\overline{J}, \overline{I}\underline{J}, \overline{I}\overline{J})] \\
    I / J                      & 0 \not\in J       & I \times J^{-1}                                                                                                    \\
    \sqrt{I}                   & I \ge 0           & [\sqrt{\overline{I}}, \sqrt{\underline{I}}]                                                                        \\
    |I|                        &                   & I \mbox{ if } I \ge 0, \quad -I \mbox{ if } I \le 0, \quad [0, \max(-\underline{I}, \overline{I})] \mbox{ otherwise} \\
    \mathcal{U}(I, J)          &                   & [\underline{I} + \underline{J} + \underline{I} \underline{J}, \overline{I} + \overline{J} + \overline{I} \overline{J}] \\
    \circ(I)                   &                   & \text{One operator is associated to each rounding mode of Table~\ref{tab/dir}}         \\
    \circe{k}(I)               &                   & \text{Several operators are associated to each rounding mode of Table~\ref{tab/dir}}         \\ \hline
  \end{array}
  $$
\end{table}

Enclosure (\texttt{BND}) is the only predicate available to users but
Gappa internally relies on more predicates to describe properties on
an expression~$x$. Such predicates appear in intermediate lemmas
of generated proofs.
$$\begin{array}{lcl}
  \mathtt{BND}(x, I) & \equiv & x \in I \\
  \mathtt{ABS}(x, I) & \equiv & |x| \in I \land I \ge 0\\
  \mathtt{FIX}(x, e) & \equiv & \exists m \in \Z, x = m \cdot 2^e\\
  \mathtt{FLT}(x, p) & \equiv & \exists m, e \in \Z, x = m \cdot 2^e \land |m| < 2^p
\end{array}
$$

The \texttt{FIX} and \texttt{FLT} predicates express that the set of
computer numbers is generally a discrete subset of the real numbers,
while intervals only consider connected subsets. They are especially
useful for automatically detecting rounded operations that actually
are exact operations, and hence do not contribute any rounding error.

Table~\ref{tab/lib} lists most of the theorems used by Gappa. The
verification process of these theorems relies on some interval
operators defined in Table~\ref{tab/int}. In particular, several
operators $\circe{k}$ related to rounding modes are needed. Some of
these operators may be left undefined; in that case, Gappa will
generate longer proofs in order to use other operators instead. Some
theorems also need to know the structure of the numbers that can be
represented with respect to a given rounding mode: $\mathbb{F}_\circ =
\{ x \in \mathbb{R}\ |\ x = \circ(x) \} = \{ m \cdot 2^e\ |\ e \ge
e_\circ \land |m| < 2^{p_\circ} \}$.

The proof script generated for Coq contains the following kind of lemma
whenever the certificate relies on interval addition to prove a
proposition, e.g.  ``if $x \in [1,2]$ (property \texttt{p1}) and $y
\in [3,4]$ (property \texttt{p2}), then $x + y \in [0,6]$ (property
\texttt{p3})''.

\begin{lstlisting}[numbers=left]
Lemma l1 : p1 -> p2 -> p3.
 intros h0 h1.
 apply add with (1 := h0) (2 := h1) ; finalize.
Qed.
\end{lstlisting}

The first line defines the lemma: if the hypotheses \texttt{p1} and
\texttt{p2} are verified, then the property \texttt{p3} is true too.
The second line starts the proof in a suitable state by using the
\texttt{intros} tactic of Coq. The third line applies the
\texttt{add} theorem of Gappa support library with the \texttt{apply}
tactic.

The \texttt{add} theorem is as follows.  \texttt{lower} and \texttt
{upper} are functions that return the lower and the upper bound of an
interval. Intervals are pairs of dyadic fractions (\texttt{FF} or $\IF$).
\texttt{Fplus2} is the addition of dyadic fractions.  \texttt{Fle2}
compares two dyadic fractions (less or equal) and returns a boolean.
The \texttt{BND} predicate holds, when its first argument, an
expression on real numbers, is an element of its second argument, an
interval defined by dyadic fraction bounds.

\begin{lstlisting}
Definition add_helper (xi yi zi : FF) :=
 Fle2 (lower zi) (Fplus2 (lower xi) (lower yi)) &&
 Fle2 (Fplus2 (upper xi) (upper yi)) (upper zi).

Theorem add :
 forall x y : R, forall xi yi zi : FF,
 BND x xi -> BND y yi ->
 add_helper xi yi zi = true ->
 BND (x + y) zi.
\end{lstlisting}

The mathematical expression of the theorem is as follows:%
$$\begin{array}{rl}
\mathtt{add}:& \forall x,y \in \R, \quad \forall I_x,I_y,I_z \in \IF,\\
&x \in I_x \Rightarrow y \in I_y \Rightarrow\\
&f_{\mathtt{add}}(I_x,I_y,I_z) = \mathit{true} \Rightarrow\\
&x + y \in I_z.\end{array}$$

If we simply needed a theorem describing the addition in interval
arithmetic, the $f_{\mathtt{add}}(I_x,I_y,I_z) = \mathit{true}$
hypothesis would be replaced by $I_x + I_y \subseteq I_z$. But we also
need for the theorem hypotheses to be automatically checkable. It is
the case for the $x \in I_x$ and $y \in I_y$ hypotheses of the
\texttt{add} theorem, since they can be directly matched to the
hypotheses \texttt{h0} ($x \in [1,2]$) and \texttt{h1} ($y \in [3,4]$)
of lemma \texttt{l1}.

Hypothesis $I_x + I_y \subseteq I_z$, however, cannot be matched so
easily. Consequently, it is replaced by an equivalent boolean
expression that can be computed by a proof checker. In lemma
\texttt{l1}, the computation is triggered by the \texttt{finalize}
tactic that checks that the current goal can be reduced to $\mathit
{true} = \mathit{true}$. This concludes the proof.

All the theorems of Gappa companion library are built the same way:
instead of having standard hypotheses that Coq would be unable to
automatically decide, they use a computable boolean expression. The
companion library formally proves that, when
this expression evaluates to \emph{true}, the standard hypotheses hold
true, and hence the goal of the theorem applies. This approach
is a simpler form of reflection techniques~\cite{Bou97}. Although the
use of booleans seems to restrict the use of Gappa to Coq proof
checker, the interval arithmetic library~\cite{DauMelMun05} developed
for PVS shows that proofs through interval computations are also
attainable to other proof assistants.

\subsection{Widening intervals to speedup proof certification}

All the interval bounds are dyadic fractions ($m \cdot 2^n$ with $m$
and $n$ relative integers) in order to ensure that the boolean
expressions are computable. Dyadic fractions are easily and
efficiently added, multiplied, and compared. Rational numbers could
also have been used: they would have been almost as efficient and
would have provided a division operator. But
common floating-point properties involved in 
certifying numerical codes are better
described and verified by using dyadic fractions.

The proof checker does not need to compute any of these dyadic numbers,
it just has to check that the interval bounds generated by Gappa make
the boolean expressions evaluate to \emph{true}, and hence are valid. 
In particular, there is absolutely no need for Gappa to compute the
sharpest enclosing interval of an expression: any wider interval can be
used. As long as the boolean expressions evaluate to \emph{true}, the
proof remains correct.

For example, manipulating the expression $x / \sqrt{3}$ will sooner or
later require $\sqrt{3} \ne 0$ to be proved. This is done by computing
an enclosing interval of $\sqrt{3}$ and verifying that its lower bound
is positive. Hence there is no need to compute an enclosing interval
with thousands of bits of precision, the interval $[1,2]$ is accurate
enough. Checking that $\sqrt{3} \in [1,2]$ holds true is fast, as it
just requires checking $1^2 \le 3 \le 2^2$. In order to get simplified
dyadic numbers in intermediate lemmas, Gappa first finds a correct proof
path and then it greedily operates backwards from the last proved
results to the first proved results, widening the intervals along the
way.

Such simplifications are important, since a proof checker like Coq is
considerably slower than a specialized mathematical library. As a
consequence, these simplified numbers can considerably speed up the
verification process of propositions, especially when they involve error
bounds. These considerations are also true for case studies: searching
for a better sub-paving and certifying it, will always be faster than
directly certifying the first sub-paving that has been found by Gappa.
The time spent by Gappa in doing all the computations over and over in
order to find a better sub-paving is negligible in comparison to the
time necessary to certify the property on one single tile with a proof
checker.

\section{Perspectives and concluding remarks}
\label{sec/conc}

In our approach to program certification, generation of proof
obligations, proof generation, and proof verification, are distinct
steps. The intermediate step is performed by Gappa with its own
computational methods, and the last one is done by a proof checker
with the help of our support library.

The developments presented so far already allowed us to guarantee the
correct behavior of many useful functions. As we continue using Gappa,
we may discover practices that cannot be handled appropriately. We
will extend Gappa, should this become necessary.  Our software, a
user's guide and details of some projects using Gappa are available on
the Internet at the address below.
\begin{center}
\url{http://lipforge.ens-lyon.fr/www/gappa/}
\end{center}

Gappa is used to certify CRlibm, a library of elementary functions
with correct rounding in the four IEEE-754 rounding modes and
performances comparable to standard mathematical libraries
\cite{DinLauMel06,DinDefLau04}.  Figure~\ref{fig/tang} presents the
input file needed to reproduce some parts of an earlier validation in
HOL Light \cite{Har97a}. These expressions define an almost correctly
rounded elementary function in single precision
\cite{Tan89}. Gappa is also used to develop robust semi-static filters
for the CGAL project \cite{MelPio07} and in the validation of delayed
linear algebra over finite fields \cite{DauGio07}.

\begin{figure}
\hrule\vspace{12pt}
\begin{verbatim}
# 1. PROG: Definitions of aliases
@rnd = float< ieee_32, ne >;

# a few floating-point constants
a1 = 8388676b-24;
a2 = 11184876b-26;
l2 = 12566158b-48;
s1 = 8572288b-23;
s2 = 13833605b-44;

# the algorithm for computing the exponential
r2 rnd= -n * l2;
r rnd= r1 + r2;
q rnd= r * r * (a1 + r * a2);
p rnd= r1 + (r2 + q);
s rnd= s1 + s2;
e rnd= s1 + (s2 + s * p);

# a few mathematical expressions to simplify later sections
R = r1 + r2;
S = s1 + s2;

E = s1 + (s2 + S * (r1 + (r2 + R * R * (a1 + R * a2))));
Er = S * (1 + R + a1 * R * R + a2 * R * R * R + 0);
E0 = S0 * (1 + R0 + a1 * R0 * R0 + a2 * R0 * R0 * R0 + Z);

# 2. PROP: Logical proposition Gappa has to verify
{ # provide the domains and accuracies of some variables
  Z in [-55b-39,55b-39] /\ S - S0 in [-1b-41,1b-41] /\
  R - R0 in [-1b-34,1b-34] /\ R in [0,0.0217] /\ n in [-10176,10176] ->
  # ask for the range of e and its absolute error
  e in ? /\ e - E0 in ? }

# 3. HINTS: Hints provided by the user
e - E0 -> (e - E) + (Er - E0);  # true as E = Er
r1 -> R - r2;                   # true as R = r1 + r2
\end{verbatim}
  \hrule
  \caption{Gappa script for proving $e$
  accurately approximates $E_0 = \exp(R_0)$ in single-precision.}
  \label{fig/tang}
\end{figure}

The whole work of generating the proof is pushed toward the external
program. All the intervals are precomputed and none of the complex
tactics of Coq are used. The proof checker only has to be able to add,
multiply, and compare integers; it does not have to be able to
manipulate rational or real numbers.  Consequently, one of our goal is
to generate proofs not only for Coq, but for other proof checkers too.

Branches and loops handling are outside the scope of this work.  Both
problems are not new to program verification and nice results have been
published in both areas.  We do not want to propose our solution for
these problems.  Our decision is to interact with the two following
tools.
\begin{itemize}
\item Why \cite{Fil03} is a tool to certify programs written in a
  generic language (C and Java can be converted to this language).  It
  certifies appropriate memory allocation and usage.  It is able to
  handle hierarchically structured code with functions and assertions.
  Why also takes care of conditional branches.  It duplicates the
  appropriate proof obligations and guarantees that both pieces of
  code meet their shared post-conditions. A floating-point formalism
  designed with Gappa in mind has recently been added to Why
  \cite{BolFil07}. Used together, Why and Gappa will be able to handle
  large pieces of software.

\item Fluctuat \cite{PutGouMar04} handles loops by effectively
  computing loop invariants. Once these invariants are provided, Gappa
  can certify the correct behavior of any numerical code. Results of
  Fluctuat will be used as oracles and certified by Gappa. Should
  there be a significant bug in Fluctuat, Gappa will stop without
  being able to meet its goals as it cannot certify erroneous results.
\end{itemize}

\bibliographystyle{acmtrans}
\bibliography{alias,perso,groupe,saao,these,livre,arith}

\end{document}